\documentclass[fleqn,usenatbib]{mnras}

\usepackage[T1]{fontenc}
\usepackage{ae,aecompl}

\usepackage{graphicx, float}	% Including figure files
\usepackage{amsmath}	% Advanced maths commands
\usepackage{amssymb}	% Extra maths symbols
\usepackage[hyphenbreaks]{breakurl}
\title[Broadband X-ray Properties of ULX M81 X--6]{Broadband X-ray Spectral Study of Ultra-luminous X-ray Source M81 X--6}%{MNRAS \LaTeXe\ template -- title goes here}

\author[Jithesh et al.]{
V. Jithesh$^{1}$\thanks{E-mail: vjithesh@iucaa.in},
C. Anjana$^{2}$ and Ranjeev Misra$^{1}$ \\
$^{1}$Inter-University Centre for Astronomy and Astrophysics (IUCAA), PB No.4, Ganeshkhind, Pune-411007, India\\
$^{2}$Department of Physics, Govt. College, Madappally, Vadakara 673102, Kerala, India\\ 
}

\date{Accepted XXX. Received YYY; in original form ZZZ}

\pubyear{2020}

\begin{document}
\label{firstpage}
\pagerange{\pageref{firstpage}--\pageref{lastpage}}
\maketitle

\begin{abstract}
Ultra-luminous X-ray sources (ULXs) are a class of extra-galactic, point-like, off-nuclear X-ray sources with X-ray luminosity from $\sim 10^{39}$~erg s$^{-1}$ to $10^{41}$~erg s$^{-1}$. We investigated the temporal and broadband X-ray spectral properties of the ULX M81~X--6 using simultaneous \textit{Suzaku} and \textit{NuSTAR} observations. To understand the nature of the source, we searched for pulsating signals from the source using the \textit{NuSTAR} observation. However, we failed to identify any strong pulsating signals from the source. Alternatively, the broadband spectral modelling with accreting magnetic neutron star continuum model provides a statistically acceptable fit, and the inferred spectral parameters and X-ray colours are consistent with other pulsating ULXs. Thus, our analysis suggests that M81~X--6 is another candidate ULX pulsar.

\end{abstract}


%%%%%%%%%%%%%%%%%%%%%%%%%%%%%%%%%%%%%%%%%%%%%%%%%%

%%%%%%%%%%%%%%%%% BODY OF PAPER %%%%%%%%%%%%%%%%%%

\section{Introduction}
%\label{sec:intro}
 
Ultra-luminous X-ray sources (ULXs) are compact, point-like, non-nuclear X-ray sources in nearby galaxies with an isotropic X-ray luminosity of $\sim 10^{39}-10^{41}$~erg s$^{-1}$ \citep[see][for a recent review]{Kaa17}. The inferred high X-ray luminosity of these sources exceeds the Eddington luminosity for a typical stellar-mass black hole (StMBH) of $10\,M_{\odot}$. Earlier studies of ULXs spectra with  the two-component phenomenological model, multi-colour disk black body \citep[MCD;][]{Mit84} plus power law (PL), suggest a cool accretion disk ($\rm kT_{in} \sim 0.1-0.3$ keV) for many sources \citep{Mil03, Mil04}. Such a cool disk interpreted as the supporting evidence for the presence of massive black holes in ULXs, which are referred to as intermediate-mass black holes (IMBHs) with a proposed mass range of $10^{2}-10^{5} M_{\odot}$ \citep{Col99, Mil04a}. Follow-up studies failed to identify any signatures of the sub-Eddington accretion flow onto the IMBH instead, the highest quality {\it XMM-Newton} observations showed peculiar features, such as soft excess (below 2 keV) and broad curvature in the 3--10 keV band \citep{Sto06}. These ubiquitous features identify ULXs in a new super-Eddington accretion state generally referred to as {\it ultra-luminous state} \citep{Gla09}. In such accretion state, ULXs can emit with super-Eddington rates and thus the majority of them are believed to be StMBH ($\rm M_{BH} < 20\,M_{\odot}$) systems.

One of the important discoveries in this field is the detection of pulsation from the ULX M82 X--2 \citep{Bac14}, which confirms the presence of neutron star (NS) compact object in the ULX population. There are five more pulsating ULXs \cite[NGC 7793 P13, NGC 5907 ULX1, NGC 300 ULX, M51 X--8 and X--7;][]{2017Sci...355..817I, 2017MNRAS.466L..48I, Car18, Bri18, Rod19} have been identified so far in external galaxies and these detections further suggest that such a high luminosity from ULXs can be even powered by an NS. These studies suggest ULXs as a heterogeneous class of objects, possibly comprises of StMBHs, IMBHs, and NS systems, and it makes them an exciting class of sources in nearby galaxies.          

%\begin{sloppypar}

Broadband X-ray spectral properties of ULXs have studied in detail using simultaneous/quasi-simultaneous \textit{Suzaku} or \textit{XMM-Newton} and \textit{NuSTAR} observations \citep{Bac13, Wal13, Ran15, Muk15, Wal15, Wal17, Wal18a}. These broadband studies with \textit{NuSTAR} demonstrate a clear cut-off at high energies for many ULXs \citep{Bac13, Wal14, Muk15}, which was marginally detected in the limited bandpass of \textit{XMM-Newton} observations. In addition, the pure thermal model description of the X-ray continuum of several ULXs resulted in a hard excess (typically above 10 keV) in the \textit{NuSTAR} data \citep{Wal13, Wal14, Muk15, Ran15, Wal15a}. Thus, the broadband X-ray spectral modelling requires an additional, high-energy spectral component to explain the observed hard excess along with thermal models. The hard excess may be arising from an optically thin, hot Comptonizing corona, while the emission from the radio jets can also describe the observed feature \citep[see also][]{Wal15a}.

Motivated by the discovery of X-ray pulsation from M82 X--2 and other sources, broadband X-ray spectral analysis of a sample of bright ULXs have been conducted to investigate the spectral similarity of this population to the known ULX pulsars \citep{Pin17, Wal18}. Adopting the commonly used spectral model for accreting magnetic NSs in the Milky Way, \citet{Pin17} described broadband spectra of most ULXs in their sample. The known ULX pulsars have the hardest spectrum among other sources in their sample, while the two ULXs, IC 342 X--1 and Holmberg IX X--1, showed the hard spectrum comparable to the known pulsating ULXs, and suggest that these two sources are candidate ULX pulsars. The phase-resolved analysis of NGC 5907 ULX \citep{Wal18} showed that the spectral form of the pulsed emission is similar to the known ULX pulsars, M82 X-2 and NGC 7793 P13. Moreover, they found that the total emission at the higher energy is dominated by the pulsed emission and successfully explained the observed hard excess. Authors further extended their study to the full sample of ULXs which have the broadband coverage and suggested that the majority of them may have an NS accretor. However, no pulsation has been detected from many of these ULXs, which may be related to the geometry of the system, where the relative contribution of the non-pulsed emission is higher than the pulsed emission. Thus, in the proposed geometry any pulsations would be diluted and harder to detect from these sources even in the high-quality spectrum \citep{Wal18}. 

%\end{sloppypar}

%\section{ULX M81 X--6}

M81~X--6 is one of the brightest non-nuclear X-ray sources in the spiral galaxy Messier 81 (M81 or NGC 3031), which is at a distance of 3.63\,Mpc \citep{Fre94}. The source is 3.4 arcmin away from the M81 AGN. M81~X--6, was discovered by the \textit{Einstein} observatory as located approximately $1'$ to the south-east of the spiral arm containing SN1993J, in the 0.2--4 keV band \citep{Fab88}. The X-ray flux of M81~X--6 from the high resolution imaging (HRI) camera and imaging proportional counter (IPC) instruments on-board \textit{Einstein} was $\sim 9.5\times10^{-13}$ and $\sim 9.8\times 10^{-13}$ erg cm$^{-2}$ s$^{-1}$, respectively. This flux has remained remarkably steady throughout its observed history with \textit{Einstein}. However, there was a marginal indication of intensity variation in the short time scale of a few minutes. In addition, the source exhibited an intensity variation by a factor of $ \sim 1.6$ in the seven {\it Advanced Satellite for Cosmology and Astrophysics} ({\it ASCA}) observations that span over three years \citep{Miz01}.

%\begin{sloppypar}
The {\it ASCA} and {\it Chandra} spectra of M81~X--6 have been described better with MCD emission from the optically-thick standard accretion disk around black hole \citep{Mak00, Miz01, Swa03}. The inferred temperature from the model fit was high ($\rm kT_{in} \sim 1.0-1.5$ keV) for a massive black hole (BH) and suggested to host a rapidly spinning BH. {\it XMM-Newton} spectrum of the source fitted with relativistic disk models such as {\tt kerrbb} and {\tt bhspec} suggests that the source can harbour a black hole with a mass range of $33-85\,M_{\odot}$ and accrete close to the Eddington limit \citep{Hui08}. Moreover, the spin value obtained from this study is consistent with a maximally spinning BH with a high inclination angle. The broadband X-ray spectral study of M81 X--6 with {\it XMM-Newton} and {\it Suzaku} observations showed the spectral downturn near $\sim 3$ keV \citep{Dew13}, which is consistent with the high-energy spectral curvature found earlier with the limited X-ray (0.3--10 keV) energy band \citep{Dew06, Sto06}. Recently, \citet{Jit18} studied the long-term spectral variability of M81~X--6 using \textit{Suzaku} and \textit{XMM-Newton} observations conducted during the period 2001--2015. They fitted the spectra mainly with the general relativistic accretion disk ({\tt kerrbb}) plus a power law component and investigated the variability. The source exhibited spectral variability which is mainly driven by the accretion rate and showed different spectral shapes during these observations. 

In this work, we make use of the availability of unpublished broadband X-ray data of the ULX M81 X--6 obtained simultaneously with \textit{NuSTAR} \citep{Har13} and \textit{Suzaku} \citep{Mit07} satellites in 2015. We carried out the temporal and spectral analysis of the source to understand the nature and broadband X-ray spectral properties. In section 2, observations and data reduction are described. The analysis and results are presented in section 3 and in section 4 we discuss and conclude our results.

\section{Observations and Data Reduction}
%\begin{sloppypar}

We used simultaneous \textit{Suzaku} (Observation ID: 710017010) and \textit{NuSTAR} (Observation ID: 60101049002) observations of the ULX M81 X--6 performed on 2015 May 18 for an exposure time of 97 and 209 ks, respectively. 

\subsection{\textit{Suzaku}}

We processed the unfiltered \textit{Suzaku} data with {\tt aepipeline} available with {\sc heasoft} version 6.22.1. We extracted the source \citep[R.A. = 09:55:32.9, decl. = +69:00:33.3, equinox J2000.0;][]{Gla09} events from a circular region of radius 80 arcsec, while the two background events were extracted from a circular source-free region of radius 110 arcsec. The high-level science products such as source and background spectra, ancillary response file, response matrix file, and light curves were generated from XIS0, XIS1 and XIS3 detectors using the standard tools available in {\sc xselect}. The XIS0 and XIS3 are the front-illuminated (FI) CCDs and their spectra were co-added using the tool {\sc addascaspec}. In the \textit{Suzaku} XIS FI spectrum, we exclude the energy range 1.7--2 keV due to calibration uncertainties. \textit{Suzaku} detectors experienced an increased charge leakage in many segments of XIS0 after 2015 March 11\footnote{\url{http://www.astro.isas.jaxa.jp/suzaku/doc/suzakumemo/suzaku_memo_2015-05.pdf}}. This anomaly resulted in the saturation of telemetry, which needs to be removed from the observed data. Since the \textit{Suzaku} observation used for this analysis is conducted after the above-mentioned date, we followed the recipe\footnote{\url{http://www.astro.isas.jaxa.jp/suzaku/analysis/xis/xis0_area_discriminaion3/Rejecting_TLMsaturation.pdf}} provided by the instrument team to remove the saturated telemetry data.

%\end{sloppypar}

%nustar analysis
\subsection{\textit{NuSTAR}}

The \textit{NuSTAR} observation was reduced with the \textit{NuSTAR} Data Analysis Software ({\sc nustardas}) Version 1.9.3. We obtained cleaned event files by applying standard corrections with the {\sc nupipeline} task and \textit{NuSTAR} CALDB version 20171002. The source and background spectra were extracted from circular regions of radius 60$''$. We analyse the high level science products with the help of {\sc heasoft} tools such as {\sc xselect} and {\sc xspec}. The source spectrum is then grouped in such a way that a bin contains at least 50 counts. All these steps are done for both focal plane modules (FPMA and FPMB) data separately. Spectra from both the modules fit jointly, without combining them.

\section{Analysis and Results}

%Light curve analysis
\subsection{Temporal Analysis}

We searched for pulsations in the \textit{NuSTAR} observation of the source. We have applied the barycentric correction to the \textit{NuSTAR} event files using the {\tt barycorr} tool available in {\sc ftools}. To search for pulsations from the source, we extracted the light curve with a bin size of 0.01\,s and performed an epoch folding \citep{Lea83} test. We employed these tasks through python by using the {\tt HENDRICS} software package \citep{2015ascl.soft02021B, 2019ApJ...881...39H}. The {\tt HENDRICS}\footnote{\url{https://hendrics.readthedocs.io/en/master/index.html}} software package, formerly called {\tt MaLTPyNT}, builds upon {\tt stingray}. It provides an accurate timing and spectral analysis of X-ray observations, especially for the \textit{NuSTAR} data. The {\tt HENDRICS} package accurately treats the data gaps in the observation due to Earth occultation or South Atlantic Anomaly and dead time in the \textit{NuSTAR} detectors. We searched for the periodic signal in the 0.01--0.1\,Hz, 0.1--1\,Hz and 1--50\,Hz frequency range using the {\tt HENefsearch} tool. However, we did not find any strong signals in these frequency ranges.  

\begin{figure}
\includegraphics[width=8.5cm, angle=0]{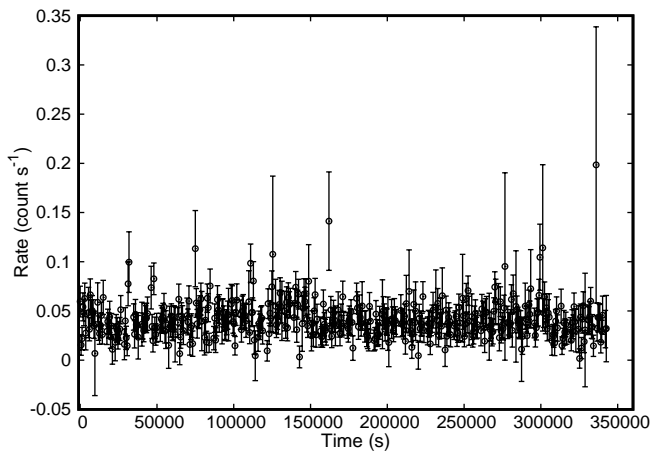}
\caption{The background subtracted \textit{NuSTAR} FPMA light curve binned with 600\,s.}
\label{lc}
\end{figure}

\begin{figure}
\includegraphics[width=8.5cm, angle=0]{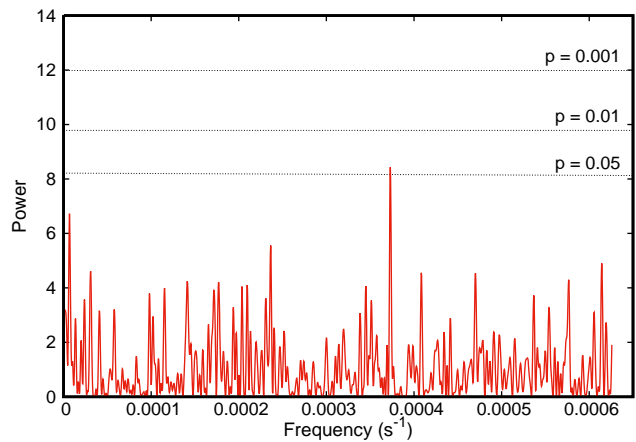}
\caption{The Generalized Lomb-Scargle periodogram for the \textit{NuSTAR} observation. The dotted lines represent different significance levels. A maximum power of 8.4 is found at $\sim 2681$\,s (just above 95 per cent level).}
\label{lsp}
\end{figure}

\begin{figure}
\includegraphics[width=8.5cm, angle=0]{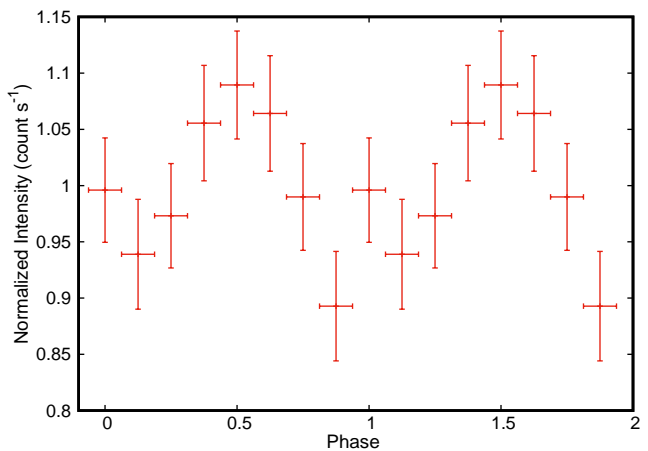}
\caption{The light curve is folded to the period of 2681\,s obtained from GLS.}
\label{fold}
\end{figure}

In order to search for the longer period modulation, we extracted the background-subtracted light curves with different bin sizes, one such light curve with a time bin size of 600\,s is shown in Figure \ref{lc}. We used the Generalised Lomb$-$Scargle periodogram \citep[GLS;][]{Zec09} and did not find any strong modulation. However, we found a maximum power of 8.4 at $\sim 2681$s (see Figure \ref{lsp}), which is not statistically significant (just above 95 per cent level). We also used the {\sc efold} tool that creates the folded light curve using the period obtained from GLS. The light curve folded to the period obtained from GLS is shown in Figure \ref{fold}.

\subsection{Broadband Spectral Analysis}

We performed broadband X-ray spectral modelling using the \textit{Suzaku} XIS FI spectrum, \textit{NuSTAR} FPMA and FPMB simultaneously in the 0.6--20 keV energy band. The background is dominating above 20 keV in the \textit{NuSTAR} spectrum. Thus, we restricted our analysis up to 20 keV energy in the \textit{NuSTAR} spectrum. The X-ray spectral fitting was performed with {\sc xspec} version 12.9.1p \citep{Arn96} and all errors are quoted at 90\% confidence level. 

The broadband spectrum initially fitted with single-component models such as multi-colour disk black body ({\tt diskbb}), power law, and cut-off power law ({\tt cutoffpl}) in {\sc xspec}. Among them, {\tt diskbb} failed to explain the observed broadband spectrum, where $\chi{^2}\rm /d.o.f=1693.2/358$. But the simple PL model improved the fit drastically compared to the {\tt diskbb} ($\Delta\chi{^2} \sim 1286$) with PL index $\Gamma=2.22\pm 0.04$ and $\chi{^2}\rm /d.o.f=407.2/358$. We incorporated the absorption in these models by two multiplicative absorption components \citep[{\tt tbabs} in {\sc xspec};][]{Wil00}. The first component describes the Galactic absorption toward the direction of the source and fixed at $N_{\rm H,Gal}=5.57\times10^{20}~\rm cm^{-2}$ \citep{Kal05}, while the second component explains the absorption local to the source and considered as a free parameter. We then replaced the simple PL model with cut-off PL which again improved the fit by $\Delta\chi{^2} \sim 67$ with one degree of freedom compared to the simple PL. The obtained cut-off energy $\sim 12.2$ keV and the photon index is $1.80^{+0.09}_{-0.10}$.
%\begin{sloppypar}

We have also attempted the two-component model, namely a combination of a disk component and a PL, for the broadband spectrum. This model provides an improved fit with $\chi{^2}\rm /d.o.f=344.9/356$, over single-component models such as {\tt diskbb} and simple PL, with $\Gamma=2.26^{+0.10}_{-0.08}$ and $\rm kT_{in}=2.07^{+0.38}_{-0.34}$ keV. Many highest quality ULXs spectra showed a soft excess and broad curvature in the limited 0.3--10 keV band \citep{Sto06, Gla09}. Such spectra are often modelled by disk plus Comptonization models. We replaced the PL with the Comptonization model, {\tt compTT}. {\tt CompTT} \citep{1994ApJ...434..570T} model describes the thermal Comptonization, where the seed photons for Comptonization are provided by the accretion disk. Thus, the inner disk temperature and seed photon temperature are tied and varied together in the analysis. The best-fit parameters yielded from this model fit are: $\rm kT_{in}=0.12^{+0.19}_{-0.04}$ keV, with $\rm kT_{e}=4.10^{+1.22}_{-0.62}$ keV and optical depth of $4.11^{+0.61}_{-0.75}$. In addition, this model provides a marginally improved fit compared to the disk plus PL model fit ($\Delta\chi{^2} \sim 4.5$) and the obtained plasma temperature is higher than the typical values obtained in the case of other ULXs \citep{Vie10, Pin12, Pin14}. 

\begin{figure}
\begin{center}
 \includegraphics[width=7.5cm,angle=0]{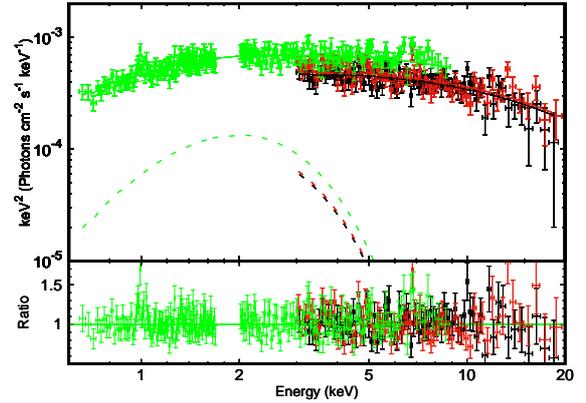}

\caption[Broadband X-ray spectrum of M81~X--6 fitted with the ``pulsator-like'' model.]
{Broadband X-ray spectrum of M81~X--6 fitted with the ``pulsator-like'' ({\tt bbody+highecut$\times$PL}) model. The {\tt bbody} and {\tt PL} components are also marked.}
\label{spectra}
\end{center}
\end{figure}

A handful of sources have been identified as pulsating ULXs (ULPs) in nearby galaxies \citep{2016ApJ...831L..14F, 2017Sci...355..817I, 2017MNRAS.466L..48I, Car18, Bri18, Rod19}. The broadband spectral properties of a large sample of ULXs, including ULPs, have been studied using simultaneous \textit{XMM-Newton} and \textit{NuSTAR} observations \citep{Pin17}. \citet{Pin17} used ``pulsator-like'' model ({\tt bbody + highecut $\times$ PL}) to describe ULPs and then compared obtained parameters with non-pulsating ULXs. The source, which we considered in this work is not included in their sample. Thus for M81~X--6, we attempted the ``pulsator-like'' model as well as the ``pulsator-like'' model without blackbody component. In this spectral modelling, there is no significant difference in the model fit between the ``pulsator-like'' model ($\chi{^2}\rm /dof=338.7/354$) and ``pulsator-like'' model without black body component ($\chi{^2}\rm /dof=339.0/356$) and yielded parameters are provided in Table \ref{Table4}. The spectral fit with the ``pulsator-like'' model is shown in Figure \ref{spectra}. We also derived the flux (using the convolution model {\tt cflux}) in the 2--4 keV, 4--6 keV and 6--30 keV (here we extended the high-energy range to 30 keV using the {\sc energies} command in {\sc xspec}) using the ``pulsator-like'' model. The \textit{hardness} is defined as the ratio of flux in the 6--30 keV and 4--6~keV band, while the \textit{softness} is defined as the ratio of flux in the 2--4 keV and 4--6 keV band. The derived \textit{hardness} and \textit{softness} values of M81 X-6 are $2.39 \pm 0.18$ and $1.81\pm 0.11$, respectively.

\begin{table*}
	\centering
%\tabletypesize{\small}
%\tablecolumns{8}
\setlength{\tabcolsep}{12.0pt}
%\tablewidth{450pt}
	\caption{Broadband X-ray Spectral modelling of M81~X--6 with the ``pulsator-like'' model. (1) Model used for spectral fitting, Model~1: {\tt bbody+highecut$\times$PL}, Model~2: {\tt highecut$\times$PL}; (2) neutral hydrogen column density in units of $10^{22}~\rm cm^{-2}$; (3) black body temperature in keV; (4) cut-off energy in keV; (5) e-folding energy in keV; (6) PL index; (7) $\chi^2$ statistics and degrees of freedom; (8) unabsorbed X-ray luminosity (0.6--20 keV band) in $10^{39} \rm erg~s^{-1}$, calculated by assuming a distance of 3.63\,Mpc~\citep{Fre94}.}
 	\begin{tabular}{cccccccc}
	\hline
	\hline
Model & $\rm N_{H}$ & kT$_{\rm bb}$ & E$_{\rm c}$ & $E_{\rm fold}$ & $\Gamma$ & $\chi{^2}$/d.o.f & $\rm L_{X}$ \\
 & ($10^{22} \rm cm^{-2}$) & (keV) & (keV) & (keV)  & & & ($10^{39}\rm erg~s^{-1}$)  \\  
\hline

Model~1 & $0.03^{+0.09}_{-0.03}$ & $0.50^{+0.33}_{-0.38}$ & $4.94^{+3.44}_{-4.94}$ & $13.17^{+11.60}_{-3.17}$ & $1.83^{+0.29}_{-0.27}$ & $338.7/354$ & $3.52^{+0.35}_{-0.22}$ \\
Model~2 & $0.07^{+0.03}_{-0.04}$ & $-$                    & $1.97^{+1.30}_{-1.97}$ & $13.19^{+4.60}_{-3.05}$ & $1.84^{+0.13}_{-0.13}$ & $339.0/356$ & $3.61^{+0.19}_{-0.19}$ \\
\hline
\end{tabular} 
\label{Table4}
%\end{deluxetable}
%\end{center}
\end{table*}

%\end{sloppypar}

\section{Discussion and Conclusion}

In this work, we performed the temporal and broadband X-ray spectral studies of ULX M81~X--6 using simultaneous \textit{Suzaku} and \textit{NuSTAR} observations in the 0.6--20 keV energy band. We investigated the temporal properties of the source using the \textit{NuSTAR} observation. To search for pulsations from M81~X--6, we used the {\tt HENDRICS} software package which properly considers the dead time of the detectors and data gaps in the observation. However, we failed to identify any strong pulsation from the source. The search for the longer period modulation provides a signal at $\sim 2681$\,s with low significance ($\sim 95\%$).

Alternatively, the pulsating nature of M81 X--6 can be investigated by the spectral modelling of the broadband spectrum. Using the broadband coverage of \textit{XMM-Newton} and \textit{NuSTAR}, \citet{Pin17} studied a sample of 12 ULXs to compare broadband X-ray spectral properties of ULPs, using the ``pulsator-like'' model, with other non-pulsating ULXs and searched potential ULP candidates. The two known ULPs (NGC 7793 P13 and NGC 5907 ULX1) have the hardest spectra \citep[\textit{hardness} $>$ 2.0 and \textit{softness} $<$ 1.5; located in the lower-right corner of Figure 2 of][]{Pin17} among the other ULXs in the colour-colour diagram, while the two non-pulsating ULXs (IC342 X--1 and Holmberg IX X--1) have X-ray colours similar to those of known ULPs. Thus, the authors suggest that IC342 X--1 and Holmberg IX X--1 are potential candidates for searching pulsations. We implemented the ``pulsator-like'' model to describe the observed broadband X-ray spectrum of M81 X--6 and yielded spectral parameters that are consistent with ULPs and potential ULP candidates. Moreover, the X-ray colours obtained for M81 X-6 are consistent with the values obtained for ULPs and ULP candidates. Thus, these results suggest that M81 X-6 is another candidate NS ULX.

\citet{Wal18} studied the full sample of ULXs to explore the spectral similarity between the known ULX pulsars and unknown ULXs using broadband observations and explained the favourable condition, i.e., the pulsed emission should dominate in the total emission, for the detection of pulsation from ULXs. In their work, they fitted broadband X-ray spectra with thick and thin accretion disk models and found that the hard excess is present in most of the ULXs. However, the pulsed emission (hard excess) is relatively less compared to the non-pulsed emission (from the accretion flow) in many ULXs they have studied. We tested this scenario by fitting the 0.6--20 keV spectrum of M81 X--6 by MCD and slim disk ({\tt diskpbb}) models, and do not see such a hard excess in high energies. Since the pulse emission typically dominated in high energies, the absence of hard excess may be the plausible explanation for the non-detection of pulsation from M81 X--6.

It is interesting to note that the earlier broadband study of M81~X--6 with \textit{Suzaku} and \textit{XMM-Newton} observations reported a cut-off at $\sim 2.8$~keV with photon index of $\sim 0.6$ and a negligible contribution at high energies in \textit{Suzaku} HXD/PIN \citep{Dew13}. However, with the broadband coverage of \textit{NuSTAR}, we observed the hard X-ray emission from M81~X--6 at least up to 20 keV and the cut-off energy yielded from the model is $\sim 12$~keV with PL index of 1.8. The obtained results are different from the previous study and suggest that the source is varied considerably between observations. This is further consistent with the long-term spectral variability of M81~X--6 \citep{Jit18}.

%Summary 
In summary, we investigated the nature and broadband properties of ULX M81 X--6 using simultaneous \textit{Suzaku} and \textit{NuSTAR} observations. We failed to identify any strong pulsating signals from the source, however, the broadband spectral modelling with accreting magnetic NS continuum model provides a statistically acceptable fit. The derived spectral parameters and X-ray colours are consistent with other pulsating ULXs and potential ULP candidates, suggest that M81~X--6 is another candidate for pulsating ULXs. The future simultaneous deep observations with \textit{XMM-Newton} and \textit{NuSTAR} can shed further light on the pulsating nature of the source.   

\section*{Acknowledgements}
We thank the referee for the constructive suggestions that improved this manuscript. AC thanks Preetha. A. U and Harikrishnan. G for the timely help, and acknowledges the IUCAA Visitors Program. VJ thanks Matteo Bachetti and Yi Xing for the help and useful discussion related to the temporal analysis. This research has made use of data obtained from the {\it Suzaku} satellite, a collaborative mission between the space agencies of Japan (JAXA) and the USA (NASA) and the {\it NuSTAR} Data Analysis Software (NUSTARDAS) jointly developed by the ASI Science Data Center (ASDC, Italy) and the California Institute of Technology (Caltech, USA). 

%%%%%%%%%%%%%%%%%%%%%%%%%%%%%%%%%%%%%%%%%%%%%%%%%%



%\bibliographystyle{mnras}
%\bibliography{ref} % if your bibtex file is called example.bib

%\begin{thebibliography}{}
%\end{thebibliography}

%%%%%%%%%%%%%%%%%%%%%%%%%%%%%%%%%%%%%%%%%%%%%%%%%%

%%%%%%%%%%%%%%%%% APPENDICES %%%%%%%%%%%%%%%%%%%%%

%\appendix



%%%%%%%%%%%%%%%%%%%%%%%%%%%%%%%%%%%%%%%%%%%%%%%%%%

% Don't change these lines
\bsp	% typesetting comment
\label{lastpage}
\end{document}